\documentclass[twocolumn]{aastex701}
\usepackage{amsmath}

\newcommand{\h}{$h_{70}^{-1}$}

\newcommand{\be}{\begin{equation}}
\newcommand{\ee}{\end{equation}}
\newcommand{\ba}{\begin{eqnarray}}
\newcommand{\ea}{\end{eqnarray}}

\newcommand{\simgt}{\lower 2pt \hbox{$\, \buildrel {\scriptstyle >}\over {\scriptstyle\sim}\,$}}
\newcommand{\simlt}{\lower 2pt \hbox{$\, \buildrel {\scriptstyle <}\over {\scriptstyle\sim}\,$}}
\newcommand{\ls}{\lower 2pt \hbox{$\;\scriptscriptstyle \buildrel<\over\sim\;$}}
\newcommand{\gs}{\lower 2pt \hbox{$\;\scriptscriptstyle \buildrel>\over\sim\;$}}

\begin{document}

\title{The Largest Catalog of Dwarf Galaxy Groups: Transient Associations or Building Blocks of the Universe?}

\author[orcid=0000-0003-3726-5611]{Marko Mi\'ci\'c}
\affil{Homer L.\ Dodge Department of Physics and Astronomy,
University of Oklahoma, Norman, OK 73019, USA}
\email[show]{micic@ou.edu}


\begin{abstract}
Elusive dwarf galaxy groups provide a rare opportunity to study hierarchical
galaxy assembly at the lowest-mass scales, but whether such
associations are gravitationally bound systems or transient
configurations remains uncertain. We present the largest catalog of isolated dwarf galaxy groups, identified using the DESI
Survey, comprising 74 systems with three to six members per group. The groups have a median line-of-sight velocity
dispersion of $\sim53$ km s$^{-1}$ and a median crossing time of
$\sim0.37$ Gyr. We find that stellar mass alone is generally
insufficient to bind the observed configurations. On the other hand, when
dark matter is included using independently inferred stellar-to-halo
mass relations and NFW potentials, 69 of 74 systems ($93\%$) favor a
bound interpretation after propagating observational and halo-model
uncertainties. For the 69 bound-favored systems,
dynamical-friction estimates indicate that substantial orbital decay
within a Hubble time is plausible if their members retain significant
dark-matter subhalos. Our results suggest that many of these groups
are systems caught during the assembly of more massive descendants,
representing some of the lowest-mass sites of ongoing hierarchical
galaxy growth in the nearby Universe and accessible laboratories for
a mode of assembly expected to have been prevalent at
early cosmic epochs.
\end{abstract}


\section{Introduction}

In the hierarchical picture of galaxy formation, present-day massive galaxies were assembled through the successive accretion and merging of smaller systems \citep[e.g.,][]{SearleZinn1978,1978MNRAS.183..341W,1991ApJ...379...52W,2014ARA&A..52..291C,BullockJohnston2005,2015MNRAS.449...49R}. At early cosmic epochs, low-mass dwarfs were a prevalent type of galaxy, dominated the merger rate per unit volume, and contributed substantially to the initial buildup of galaxy mass \citep[e.g.,][]{Klypin1999,Moore1999,2015ApJ...813...21M,2024Natur.626..975A}. However, despite recent technological advances, direct observations of hierarchical assembly at high redshift remain challenging. Nearby interacting dwarfs therefore provide an accessible window into this process, offering a unique opportunity to study the early stages of galaxy growth in detail in the local Universe.

Dwarf galaxy groups, composed exclusively of dwarf galaxies, are particularly valuable laboratories for this purpose \citep[e.g,][]{Simon2019}. While various works compiled catalogs of local dwarf--dwarf pairs \citep{2015ApJ...805....2S,2018ApJS..237...36P,2020AJ....159..103K,2024ApJ...968L..21M,2024MNRAS.533.3771L,2025A&A...702A.222C}, dwarf galaxy groups remain exceptionally scarce. Only seven such spectroscopically confirmed systems, consisting of three to five members, are known to date \citep{2017NatAs...1E..25S}. Unlike dwarf satellites orbiting a massive galaxy, isolated dwarf groups allow the effects of dwarf--dwarf interactions and their common gravitational environment to be studied without the dominant tidal field of a massive host. If gravitationally bound, these systems may represent hierarchical structures caught at an early stage of assembly: multiple low-mass galaxies sharing a dark-matter environment and evolving via concurrent and consecutive interactions into a more massive descendant. Their internal dynamics and eventual coalescence may therefore provide an insight into one of the fundamental pathways through which galaxies were assembled in the early Universe.

Here, we present the second study of the Dark Energy Survey Instrument (DESI) Interacting Dwarfs Survey \citep{MicicKaviraj2026}. Using spectroscopic redshifts and galaxy properties from DESI, we identified 74 isolated dwarf systems consisting of three to six members. We characterize their projected structure, stellar content, internal kinematics, crossing times, dynamical masses, and dark-matter halo masses to address a fundamental question: are dwarf galaxy groups bona fide early-Universe analogs, gravitationally bound structures representing the building blocks of the hierarchical assembly of galaxies, or transient associations observed during brief encounters?

\section{Sample Selection}
\subsection{DESI Survey}
DESI is a wide-area spectroscopic survey designed to map the large-scale structure of the Universe by inferring redshifts of more than 30 million galaxies \citep{2016arXiv161100036D}. The Early Data Release (EDR) provides the first publicly available spectroscopic dataset from DESI \citep{2024AJ....168...58D}. For our analysis, we used the value-added DESI EDR catalog, which derives physical properties of more than a million galaxies \citep{2024A&A...691A.308S}, and can be accessed through the VizieR Catalogue Service \citep{2024yCat..36910308S}.

The stellar masses are derived with CIGALE \citep{Boquien2019} from Legacy Surveys $grz$ and WISE photometry, assuming a \citet{Chabrier2003} initial mass function and solar metallicity. Dwarf galaxies are typically metal-poor \citep{Kirby2013}, but \citet{2024A&A...691A.308S} find that these assumptions change the stellar masses by at most $\sim0.25$ dex, comparable to the stellar mass uncertainty adopted in Section \ref{stat}.
\subsection{Selection Criteria}
We select dwarf galaxy groups by requiring galaxies to simultaneously meet the stellar mass, redshift, projected separation, and environment criteria. Candidate group members must: (1) have stellar mass of $\log(M_\star/M_\odot)<9.5$; (2) lie within $|\Delta z|\leq0.001$ of the median group redshift, corresponding to a rest-frame line-of-sight velocity offset of approximately $|\Delta v|\lesssim300~{\rm km\,s^{-1}}$; (3) lie within a projected distance of $100$ kpc from the stellar-mass-weighted group centroid, ensuring that the selected systems are spatially compact; (4) have no non-dwarf neighbors within a projected distance of $1$ Mpc. The systems containing at least three dwarfs that simultaneously satisfy these requirements are labeled as groups.
\subsection{Sample}
Our final sample consists of 74 dwarf galaxy groups containing 238 galaxies. Low-richness systems dominate the sample: 63 groups contain three members, seven contain four members, three contain five members, and one contains six members. The groups span redshifts of $0.02<z<0.33$, with a median of $z=0.07$. Individual galaxy stellar masses range from $\log(M_\star/M_\odot)=5.7$ to $9.5$, while the total stellar masses of the groups span $\log(M_{\star,\mathrm{group}}/M_\odot)=7.9$--$9.8$, with a median of $9.3$. The sample encompasses a broad range of projected sizes and configurations, with the most compact system being a triplet with a maximum projected separation between its members of only $2.6$ kpc. The catalogs of the groups and of their member galaxies are described in Appendix \ref{app:catalog}.

\section{Dynamical Nature and Evolution of Dwarf Galaxy Groups}
The compact galactic associations identified here do not necessarily need to represent gravitationally bound structures. Instead, they may be transient associations or chance alignments. We examine their internal dynamics to establish their true physical nature and dynamical state.
\subsection{Internal Kinematics and Dynamical Properties}
For each system, we define the median redshift of its members as the group redshift, $z_{\rm grp}$, and calculate the line-of-sight (LOS) velocity offset of each member as:
\begin{equation}
\Delta v_i =
c\frac{z_i-z_{\rm grp}}{1+z_{\rm grp}}.
\end{equation}
The resulting velocity offsets range from $-285$ to $+269~{\rm km\,s^{-1}}$. Next, we calculate the LOS velocity dispersion, $\sigma_v$, of each group. Because the groups contain only three to six members, we adopt the gapper estimator of the velocity dispersion \citep{1990AJ....100...32B}. The resulting dispersions span 8.4 $\lesssim \sigma_v \lesssim$ 228 $\rm$ km s$^{-1}$
with a median of $\sigma_v\simeq53$ km s$^{-1}$.

\begin{figure}[ht!]
\includegraphics[width=\columnwidth]{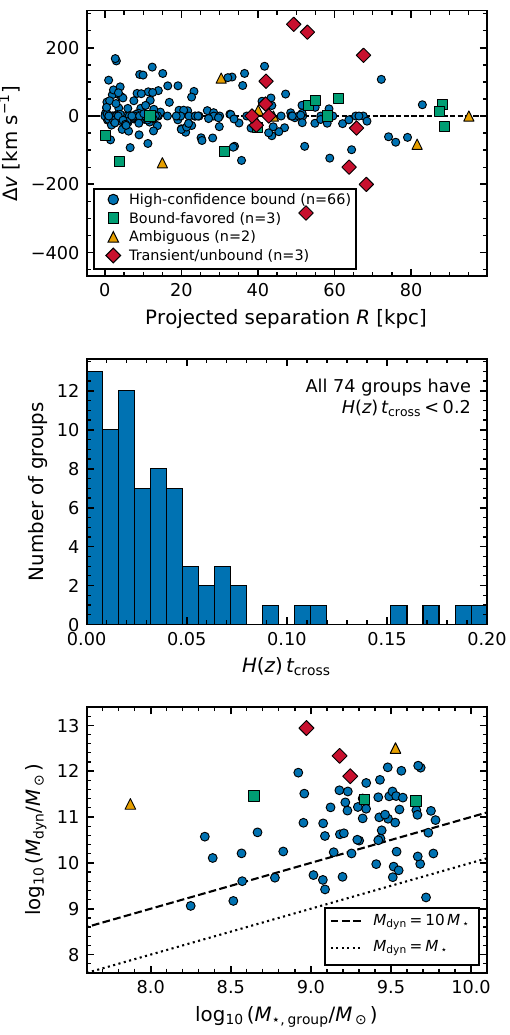}
	\caption{Top: Line-of-sight velocity offset from the group median versus projected distance from the group's mass-weighted centroid, with points colored by the four-tier Monte Carlo bound classification
(Section \ref{stat}). Middle: Distribution of
crossing times relative to the Hubble time at each group's redshift. Bottom: Group dynamical mass versus total stellar mass, with the same colors and marker shapes as the top panel. The dotted and dashed lines mark $M_{\rm dyn}=M_\star$ and $M_{\rm dyn}=10\,M_\star$, respectively.
\label{fig1}}
\end{figure}

We next characterize the projected physical sizes of the groups. For each system, we determine the stellar-mass-weighted centroid, and then infer the size of the system using the stellar-mass-weighted rms radius given as:
\begin{equation}
R_{\rm rms}
=
\left[
\frac{\sum_i M_{\star,i}R_i^2}
{\sum_i M_{\star,i}}
\right]^{1/2},
\end{equation}
where $M_{\star,i}$ and $R_i$ are stellar mass and projected distance from the stellar-mass-weighted centroid of the $i$th member. The groups span $R_{\rm rms}\simeq0.2$--$86$ kpc, with a median of $\simeq18$ kpc. The projected LOS velocity offsets vs. projected distances from the group centers are shown in the top panel of Figure \ref{fig1}.

The combination of the characteristic physical size and internal velocity scale provides an estimate of the time required for a galaxy to traverse the system:
\begin{equation}
t_{\rm cross}
=
\frac{R_{\rm rms}}{\sigma_v}.
\end{equation}
The resulting crossing times span 0.005 $\lesssim$ t$_{\rm cross} \lesssim$ 2.620 ${\rm Gyr}$,
with a median of $t_{\rm cross}\simeq0.37$ Gyr. To place these crossing times on a physically meaningful scale, we compare them with the Hubble time at the redshift of each group through the dimensionless quantity H(z)t$_{\rm cross}$. Here $H(z)=H_0\sqrt{\Omega_m(1+z)^3+\Omega_\Lambda}$, and we adopt a flat $\Lambda$CDM cosmology with $H_0=70$ km s$^{-1}$ Mpc$^{-1}$, $\Omega_m=0.3$, and $\Omega_\Lambda=0.7$ throughout. All 74 groups have crossing times shorter than the Hubble time ($H(z)t_{\rm cross}<1$), shown in the middle panel of Figure \ref{fig1}. Thus, if gravitationally bound, the groups have sufficient time to undergo substantial dynamical evolution within a Hubble time.

Finally, we estimate the characteristic dynamical mass for each group. We adopt the projected virial mass estimator \citep{BahcallTremaine1981,1985ApJ...298....8H}:
\begin{equation}
M_{\rm dyn}
=
\frac{3\pi}{2G}\sigma_v^2 R_{\rm PV},
\end{equation}
where R$_{\rm PV}$ is the projected virial radius, calculated from the projected pairwise separations. The inferred dynamical masses are typically higher than the observed stellar masses of the groups, as expected if the systems reside within dark-matter-dominated potentials, shown in the bottom panel of Figure \ref{fig1}.
\subsection{Can the Stellar Mass Bind the Groups?}
\label{stellarbind}
The first question we address is whether the observed stellar mass is sufficient to account for the internal dynamics of the groups. For each system, we compute the projected stellar kinetic and potential energies:
\begin{equation}
T_{\rm LOS,\star}
=
\frac{1}{2}\sum_i M_{\star,i}(\Delta v_i)^2,
\end{equation}
and
\begin{equation}
U_{\rm proj,\star}
=
-G\sum_{i<j}
\frac{M_{\star,i}M_{\star,j}}{R_{ij}},
\end{equation}
where $R_{ij}$ is the projected separation between galaxies $i$ and $j$. We then define the projected stellar virial ratio as:
\begin{equation}
Q_{\star}
=
\frac{2T_{\rm LOS,\star}}
{|U_{\rm proj,\star}|}.
\end{equation}
Systems with $Q_{\star}<2$ have negative projected stellar energy and are therefore consistent with being bound by their stellar mass alone \citep{BertoldiMcKee1992}. However, only 15/74 groups ($20\%$) satisfy this condition, while the median value is $Q_{\star}\simeq14.2$ (Figure \ref{fig4}, top panel). For most systems, the observed stellar component alone is insufficient
to gravitationally bind the galaxies at their measured relative
velocities.

As a complementary test, we apply the Newtonian two-body binding
criterion of \citet{1982ApJ...257...23B} to every galaxy pair. A
necessary condition for a bound solution at any projection angle is:
\begin{equation}
V_r^2R_p
\leq
\frac{4}{3\sqrt{3}}GM_{\star,\rm pair},
\end{equation}
where $V_r$, $R_p$, and $M_{\star,\rm pair}$ are the relative LOS
velocity, projected separation, and total stellar mass of the pair,
respectively. Only 1/74 groups has all galaxy pairs
satisfying this criterion. Both tests indicate that stellar mass alone is generally insufficient to bind the observed configurations.
\subsection{Can the Dark Matter Bind the Groups?}
\label{dmbind}
The stellar-only tests above do not take into account the dominant gravitating component. We therefore test whether the observed group
configurations can be supported by independently inferred dark-matter
halos. We estimate a characteristic halo mass, $M_{200}$, for each
group by applying the redshift-dependent stellar-to-halo mass relation
(SHMR) of \citet{2013MNRAS.428.3121M} to the total stellar mass of the
system:
\begin{equation}
\frac{M_{\star,\rm group}}{M_{200}}
=
2N
\left[
\left(\frac{M_{200}}{M_1}\right)^{-\beta}
+
\left(\frac{M_{200}}{M_1}\right)^{\gamma}
\right]^{-1},
\end{equation}
where $M_1$ is the characteristic halo mass, $N$ is the normalization, and $\beta$ and $\gamma$ are the low- and high-mass slopes. These parameters evolve with redshift following
\citet{2013MNRAS.428.3121M}:
\begin{eqnarray}
\log_{10}(M_1/M_\odot) &=& 11.590+1.195\,\frac{z}{1+z}, \nonumber\\
N &=& 0.0351-0.0247\,\frac{z}{1+z}, \nonumber\\
\beta &=& 1.376-0.826\,\frac{z}{1+z}, \nonumber\\
\gamma &=& 0.608+0.329\,\frac{z}{1+z}.
\end{eqnarray}
For each inferred halo mass, we calculate $R_{200}$, defined as the
radius within which the mean enclosed density is 200 times the
critical density of the Universe at the group redshift, and adopt an NFW
density profile \citep{1997ApJ...490..493N}. Halo concentrations are
assigned using the redshift-dependent mass-concentration relation of
\citet{2014MNRAS.441.3359D}:
\begin{equation}
\log_{10}c_{200}=a+b\,\log_{10}\left(\frac{M_{200}}{10^{12}\,h^{-1}M_\odot}\right),
\end{equation}
with $a=0.520+0.385\exp(-0.617\,z^{1.21})$, $b=-0.101+0.026\,z$, and $h=0.7$. The corresponding escape velocity at
radius $r$ is:
\begin{equation}
v_{\rm esc,NFW}(r)
=
\left[
\frac{2GM_{200}}{r\,f(c_{200})}
\ln\left(1+\frac{r}{r_s}\right)
\right]^{1/2},
\end{equation}
where $c_{200}$ is the halo concentration, $r_s=R_{200}/c_{200}$ is
the NFW scale radius, and
$f(c_{200})=\ln(1+c_{200})-c_{200}/(1+c_{200})$. A system is considered bound-consistent if all observed members lie
within $R_{200}$ and their LOS velocity offsets satisfy
$|\Delta v_i|<v_{\rm esc,NFW}(R_i)$ \citep{2016A&A...588A..14T}.

\begin{figure}[ht!]
\includegraphics[width=\columnwidth]{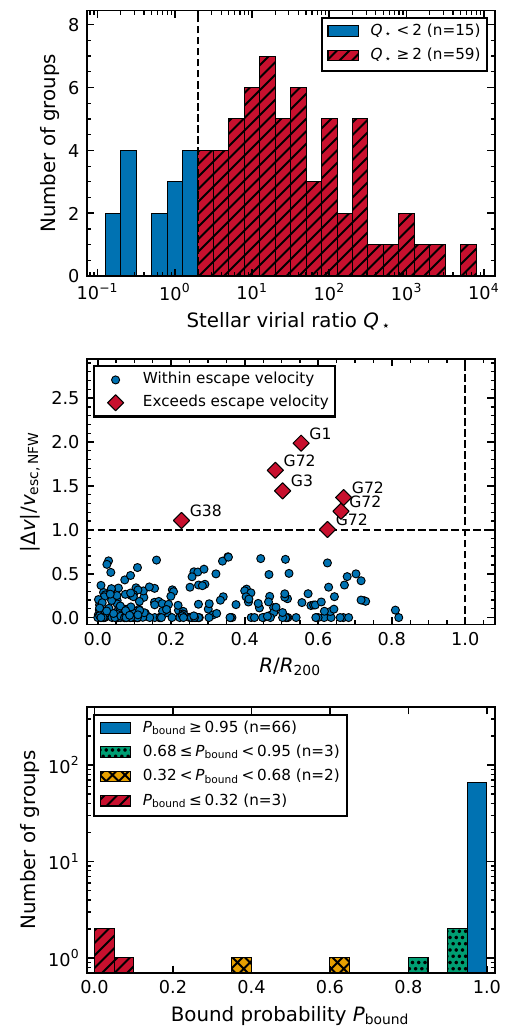}
	\caption{Three tests of gravitational boundedness. Top: Stellar-mass-only virial ratio $Q_\star = 2T_{\rm LOS,\star}/|U_{\rm proj,\star}|$; only 15/74 groups (blue, unhatched) satisfy $Q_\star<2$ (vertical dashed line). Middle: Per-member velocity offsets normalized by the NFW-based escape velocity. Once halo mass is included, all but 7 members (4 groups; red diamonds labeled by group number) fall within the bound envelope. Bottom: Monte Carlo-propagated bound probability $P_{\rm bound}$, incorporating velocity, stellar mass, and SHMR/concentration uncertainties; 66/74 groups (89\%) are robustly bound at $\geq$95\% confidence. The vertical axis of this panel is logarithmic so that the sparsely populated classes remain visible.
	\label{fig4}}
\end{figure}

Overall, 70 of 74 groups ($94.6\%$) satisfy both criteria and are consistent with being gravitationally bound within their dark-matter halos (Figure \ref{fig4}, middle panel). This contrasts with the
stellar-only analysis, where we found only 15/74 bound systems, demonstrating the critical
role of dark matter in the dynamics of these systems.
\subsection{Statistical Assessment of Group Boundness}
\label{stat}
While the criteria above provide an initial assessment of the
groups boundness, projection effects and uncertainties
in the derived halo properties can affect the resulting
classifications. We therefore quantify the robustness using 5000
Monte Carlo realizations for each group. In each realization, we
perturb the member velocities and stellar masses assuming uncertainties
of $10$ km s$^{-1}$ and $0.25$ dex, respectively \citep{2023ApJ...943...68L,2024A&A...691A.308S}, and propagate the
uncertainties and intrinsic scatter in the SHMR
\citep{2013MNRAS.428.3121M} and halo mass-concentration relation
\citep{2014MNRAS.441.3359D}. We define $P_{\rm bound}$ as the fraction
of realizations in which all members lie within $R_{200}$ and satisfy
$|\Delta v_i|<v_{\rm esc,NFW}(R_i)$.

The resulting $P_{\rm bound}$ distribution is shown in the bottom panel of 
Figure~\ref{fig4}. Out of the 74 groups, 66
($89\%$) have $P_{\rm bound}\geq0.95$ and are classified as
robustly bound systems. Three systems ($4\%$) are
bound-favored, with $0.68\leq P_{\rm bound}<0.95$, while two
($3\%$) are ambiguous, with $0.32<P_{\rm bound}<0.68$. The remaining
three systems ($4\%$) favor a transient or unbound interpretation,
with $P_{\rm bound}\leq0.32$. The sample is therefore dominated by systems whose observed configurations are robustly consistent with gravitationally bound
dwarf groups.
Finally, we test whether these results depend on the adopted group definition by repeating the full analysis with stricter selection thresholds. Restricting the redshift window to $|\Delta z|\leq0.0005$, the projected radius to 50 kpc, or the stellar mass limit to $\log(M_\star/M_\odot)<9.0$ leaves 68, 44, and 24 groups, respectively. In these three cases, 12\%--25\% of the groups can be bound by their stellar mass alone and 88\%--98\% have $P_{\rm bound}\geq0.68$, compared with 20\% and 93\% for the fiducial sample. Applying all three stricter thresholds simultaneously leaves 16 groups, of which 2 can be bound by their stellar mass alone and 15 (94\%) have $P_{\rm bound}\geq0.68$. We do not consider looser thresholds because the fiducial limits already match the physical scales of the inferred halos: the 100 kpc limit is comparable to the median virial radius ($R_{200}\simeq108$ kpc), and the velocity window ($\sim225$--$295$ km s$^{-1}$) is comparable to the largest escape velocities ($\lesssim350$ km s$^{-1}$). Members admitted by looser limits would therefore lie outside the virial radius or exceed the escape velocity, and could not be bound to the group.
\subsection{The Fate of Bound Dwarf Groups}
\label{sec:fate}
The short crossing times of the bound-consistent groups indicate that
their members can undergo substantial dynamical evolution within a
Hubble time, but the crossing time does not represent a merger
timescale. We estimate the orbital-decay timescale of each group member except the most massive galaxy, which we adopt as the central galaxy, using the Chandrasekhar dynamical-friction
approximation:
\begin{equation}
t_{\rm df} =
\frac{1.17}{\ln\Lambda}
\frac{r^2 V_c(r)}{Gm_{\rm sat}},
\end{equation}
where $V_c(r)$ is calculated from the independently inferred NFW group
halo,
$\ln\Lambda=\ln(1+M_{200}/m_{\rm sat})$ \citep{Chandrasekhar1943, LaceyCole1993}, and $m_{\rm sat}$ is the total bound mass of the satellite. We restrict this calculation
to the 69 systems with $P_{\rm bound}\geq0.68$. For each group, we use the largest
$t_{\rm df}$ among its members, $t_{\rm df,max}$, as a conservative
estimate of whether all currently identified members can sink within
a Hubble time.
\begin{figure}[ht!]
\includegraphics[width=\columnwidth]{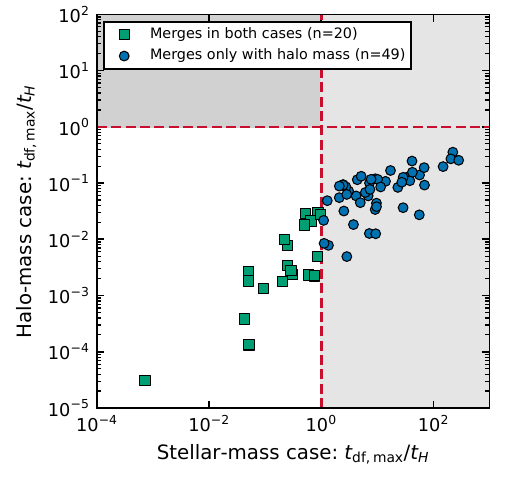}
	\caption{Merger timescale for the 69 bound-consistent groups ($P_{\rm bound}\geq0.68$), comparing the stellar-mass-only and SHMR-based halo-mass dynamical friction estimates. For each group, the plotted timescale is that of its slowest-sinking satellite, normalized by the Hubble time at the group's redshift. Under the halo-mass assumption, all 69 groups (100\%) fall below $t_{\rm merge}/t_H=1$ (horizontal red dashed line), indicating each is expected to coalesce into a single galaxy within a Hubble time. Under the stellar-mass-only assumption, only 20/69 groups (green squares) satisfy this criterion (vertical red dashed line).
	\label{figg3}}
\end{figure}
The primary uncertainty here is the bound
satellite mass. Dwarf galaxies are expected to retain dark matter, but
their halos may have experienced substantial tidal stripping after
entering the group potential \citep{Deason2014,Wetzel2015}. For this reason, we adopt two limiting assumptions. In the most conservative case, we set
$m_{\rm sat}=M_{\star}$, while in the halo-based case we adopt the
individual SHMR-inferred halo mass of each satellite. We adopt the projected group-centric separation,
$r=R_{\rm proj}$, as the characteristic orbital radius.

Figure~\ref{figg3} compares $t_{\rm df,max}/t_H$ for
the two satellite-mass assumptions. Even in the stellar-only limiting
case, 20 of 69 groups ($29\%$) have $t_{\rm df,max}<t_H$. When the
SHMR-inferred satellite halo masses are included, all 69 groups
satisfy $t_{\rm df,max}<t_H$. However, these estimates should be taken with caution. The Chandrasekhar approximation does not capture the full orbital
evolution of interacting galaxies, and the amount of dark matter
retained by satellites is uncertain. Nevertheless, the
calculation demonstrates that, if the dwarfs retain substantial
dark-matter subhalos, dynamical friction is sufficient for
the dwarf systems to undergo orbital decay within a Hubble time plausibly.
\section{Conclusions}
\label{sec:conclusions}

We present the largest catalog to date of isolated dwarf galaxy
groups, comprising 74 systems and 238 galaxies identified using the DESI
Survey. The groups contain three to six members and have short
dynamical timescales, with median $\sigma_v\simeq53$ km s$^{-1}$ and
$t_{\rm cross}\simeq0.37$ Gyr. We find that stellar mass alone is generally insufficient to bind the
observed configurations. However, inclusion of the dark-matter halos produces a different picture.
After propagating observational and halo-model uncertainties, 69 of
74 systems ($93\%$) favor a bound interpretation with
$P_{\rm bound}\geq0.68$, including 66 ($89\%$) with
$P_{\rm bound}\geq0.95$. For these 69 systems, dynamical-friction
estimates indicate that substantial orbital decay within a
Hubble time is plausible if their members retain significant
dark-matter subhalos. These results suggest that many of the dwarf groups
are systems caught during the assembly of a more massive
galaxy. They may therefore
represent some of the lowest-mass sites of ongoing hierarchical galaxy
growth in the nearby Universe, and surviving examples of a mode of
assembly expected to have been considerably more common at early
cosmic epochs.

\begin{acknowledgments}
M.M. is supported by the Dodge Family Postdoctoral Fellowship at the University of Oklahoma. M.M. thanks the anonymous reviewer for comments that improved this manuscript, and Xinyu Dai and Sean Matt for reading the manuscript and for constructive discussions.
\end{acknowledgments}



\appendix
\setcounter{table}{0}\renewcommand{\thetable}{A\arabic{table}}
\section{Dwarf Group Catalog Description}
\label{app:catalog}
Tables \ref{tab:groups} and \ref{tab:members} provide a description of the catalogs of the 74 dwarf galaxy groups and their 238 member galaxies derived in this work. The complete datasets are available in machine-readable format. Values are reported to a precision consistent with the adopted uncertainties of 10 km s$^{-1}$ in velocity and 0.25 dex in stellar mass (Section \ref{stat}).
\begin{deluxetable}{lll}
\tablecaption{Description of the Dwarf Group Catalog\label{tab:groups}}
\tablehead{\colhead{Label} & \colhead{Unit} & \colhead{Description}}
\startdata
Group & \nodata & Group identifier \\
N & \nodata & Number of member galaxies \\
RAdeg & deg & R.A. of group centroid, decimal degrees (J2000) \\
DEdeg & deg & Decl. of group centroid, decimal degrees (J2000) \\
z & \nodata & Median redshift of the members \\
logMstar & log($M_\odot$) & log, total stellar mass of the group, solar units \\
sigma\_v & km s$^{-1}$ & Line-of-sight velocity dispersion \\
Rrms & kpc & Mass-weighted rms projected radius \\
tcross & Gyr & Crossing time \\
logMdyn & log($M_\odot$) & log, dynamical mass, solar units \\
logM200 & log($M_\odot$) & log, SHMR-inferred halo mass, solar units \\
Pbound & \nodata & Monte Carlo bound probability \\
\enddata
\tablecomments{Group centroids are stellar-mass weighted. Redshifts and stellar masses are taken from M. Siudek et al. (2024). This table is available in its entirety in machine-readable form.}
\end{deluxetable}
\begin{deluxetable}{lll}
\tablecaption{Description of the Member Galaxy Catalog\label{tab:members}}
\tablehead{\colhead{Label} & \colhead{Unit} & \colhead{Description}}
\startdata
Galaxy & \nodata & Galaxy identifier (group and member number) \\
Group & \nodata & Host group identifier \\
RAdeg & deg & R.A., decimal degrees (J2000) \\
DEdeg & deg & Decl., decimal degrees (J2000) \\
z & \nodata & DESI spectroscopic redshift \\
logMstar & log($M_\odot$) & log, total stellar mass, solar units \\
Delta\_v & km s$^{-1}$ & Velocity offset from the group median \\
R & kpc & Projected distance from the group centroid \\
\enddata
\tablecomments{Members of each group are listed on consecutive rows. Coordinates, redshifts, and stellar masses are taken from M. Siudek et al. (2024). This table is available in its entirety in machine-readable form.}
\end{deluxetable}

\bibliography{dids}{}
\bibliographystyle{aasjournalv7}

\end{document}